\documentclass[twocolumn,showpacs,preprintnumbers,amsmath,amssymb]{revtex4}

\usepackage{graphicx}
\usepackage{graphics}
\usepackage{dcolumn}
\usepackage{bm}
\usepackage{color}

\begin{document}

\preprint{APS/123-QED}
\bibliographystyle{prsty}
\title{Evidence for an excitonic insulator phase in 1\textit{T}-TiSe$_{2}$}
\author{H. Cercellier}
 \email{herve.cercellier@unine.ch}
\author{C. Monney}
\author{F. Clerc}
\author{C. Battaglia}
\author{L. Despont}
\author{M. G. Garnier}
\author{H. Beck}
\author{P. Aebi}%

\affiliation{%
Institut de Physique, Universit\'e de Neuch\^atel, CH-2000 Neuch\^atel, Switzerland}%

\author{L. Patthey}
\affiliation{Swiss Light Source, Paul Scherrer Institute, CH-5232 Villigen, Switzerland
}%

\author{H. Berger}
\affiliation{Institut de Physique de la Mati\`ere Complexe, EPFL, CH-1015 Lausanne, Switzerland
}%

\date{\today}

\begin{abstract}

We present a new high-resolution angle-resolved photoemission 
study of 1\textit{T}-TiSe$_{2}$ in both, its
room-temperature, normal phase and its low-temperature,
charge-density wave phase. At low temperature the
photoemission spectra are strongly modified, with large band
renormalisations at high-symmetry points of the Brillouin zone
and a very large transfer of spectral weight to backfolded bands.
A theoretical calculation of the spectral function for an excitonic
insulator phase reproduces the experimental features with very
good agreement. This gives strong evidence in favour of
the excitonic insulator scenario as a driving 
force for the charge-density wave transition in 1\textit{T}-TiSe$_{2}$.
\end{abstract}


\maketitle{}

Transition-metal dichalcogenides (TMDC's) are layered compounds
exhibiting a variety of interesting physical properties, mainly due to
their reduced dimensionality \cite{AdvPhys24117}. 
One of the most frequent
characteristics is a ground state exhibiting a charge-density wave
(CDW), with its origin arising from a particular
topology of the Fermi surface and/or a strong electron-phonon
coupling \cite{clerc:155114}. 
Among the TMDC's 1\textit{T}-TiSe$_{2}$ shows a commensurate 2$\times$2$\times$2 
structural distortion below 202 K, accompanied by the softening of a zone boundary phonon
and with changes in the transport properties
\cite{PhysRevB.14.4321,PhysRevLett.86.3799}. In spite of many
experimental and theoretical studies, the driving force for the
transition remains controversial. Several angle-resolved
photoelectron spectroscopy (ARPES) studies suggested either the onset of an
excitonic insulator phase \cite{PhysRevB.61.16213,PhysRevLett.88.226402} or a band Jahn-Teller effect \cite{PhysRevB.65.235101}. Furthermore, TiSe$_{2}$ has recently attracted strong interest due to the observation of superconductivity when intercalated with Cu \cite{Morosan2006}. 
In systems showing exotic properties, such as Kondo systems
for example \cite{Adv_Phys_45_299_1996_Malterre}, the calculation of the spectral function has often
been a necessary and decisive step for the interpretation of the
ARPES data and the determination of the ground state of the
systems. In the case of 1\textit{T}-TiSe$_{2}$, such a calculation for an excitonic insulator phase lacked so far.  
 
 In this letter we present a high-resolution ARPES study of
 1\textit{T}-TiSe$_{2}$, together with theoretical calculations of the
 excitonic insulator phase spectral function for this compound.
We find that the experimental ARPES spectra show strong band renormalisations with a very large transfer of spectral weight into backfolded bands in the low-temperature phase.  The spectral function calculated for the excitonic insulator phase is in strikingly good agreement with the experiments, giving strong evidence for the excitonic origin of the transition.

\begin{figure}
\begin{center}
\scalebox{0.48}{\includegraphics{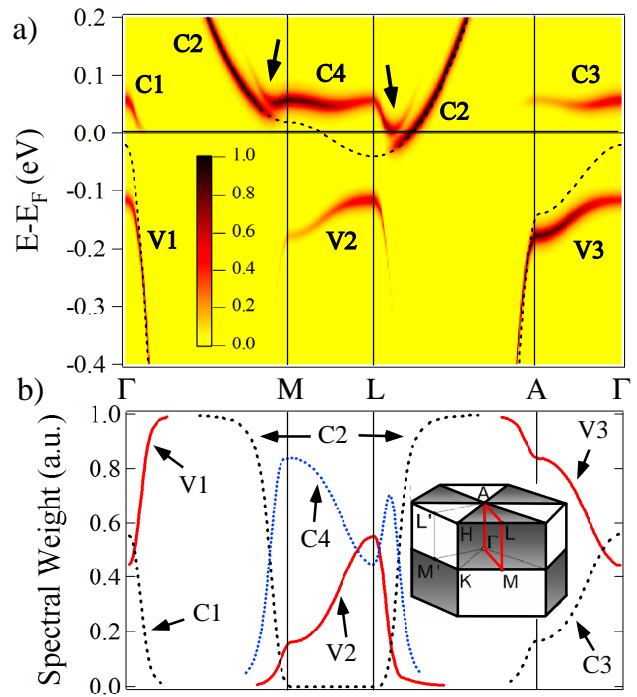}}
\caption{\label{fig1}: a) Spectral function of the excitonic
insulator in a 1\textit{T} structure calculated for a 20 meV
overlap and an order parameter $\Delta$=0.05 eV. The V1-V3 (resp. C1-C4) branches refer to the valence (resp. conduction) band. Dashed lines
correspond to the normal phase ($\Delta$=0). The path in
reciprocal space is shown in red in the inset. b) Spectral weight
of the different bands. Inset : bulk Brillouin zone of 1$T$-TiSe$_{2}$.}
\end{center}
\end{figure}

The excitonic insulator model was
first introduced in the sixties, for a semi-conductor or a
semi-metal with a very small indirect gap E$_{G}$
\cite{PhysRevLett.19.439,RevModPhys.40.755,PhysRev.158.462,bronold:165107}. Thermal excitations lead to the formation of holes in the valence band and electrons
in the conduction band. For low free carrier densities, the weak screening of the
electron-hole Coulomb interaction leads to the formation of
stable electron-hole bound states, called excitons. If the exciton binding energy E$_{B}$ is larger than the
gap energy E$_{G}$, the system becomes unstable upon formation of
excitons. This instability can drive a transition to a
coherent ground state of condensed excitons, with a periodicity
given by the spanning vector $\textbf{w}$ that connects the valence band maximum to the conduction band minimum. 
In the particular case
of TiSe$_{2}$, there are three vectors ($\textbf{w}_{i}$, $i=1,2,3$) connecting
the Se 4p-derived valence band maximum at the $\Gamma$ point to the three
symmetry-equivalent Ti 3d-derived conduction band minima at the $L$ points of the
Brillouin zone (BZ) (see inset of fig. \ref{fig1}b)).


Our calculations are based on the BCS-like model of J\'erome, Rice
and Kohn \cite{PhysRev.158.462}, adapted for multiple $\textbf{w}_{i}$.   
The band dispersions for the normal phase have been chosen of the form
\begin{eqnarray}\label{dispersions}
\epsilon_{v}(\textbf{k})=\epsilon_{v}^{0}+\hbar^{2}\frac{k_{x}^{2}+k_{y}^{2}}{2m_{v}}+t_{v}\cos(\frac{2\pi k_{z}}{c})\nonumber
\\
\epsilon_{c}^{i}(\textbf{k},\textbf{w}_{i})=\epsilon_{c}^{0}+\hbar^{2}\Big(\frac{(k_{x}-w_{ix})^{2}}{2m_{c}^{x}}+\frac{(k_{y}-w_{iy})^{2}}{2m_{c}^{y}}\Big)\nonumber\\
+t_{c}\cos\Big(\frac{2\pi(k_{z}-w_{iz})}{c}\Big)
\end{eqnarray}
for the valence ($\epsilon_{v}$) and the three conduction ($\epsilon_{c}^{i}$) bands respectively, 
with $c$  the lattice parameter perpendicular to the surface in the normal (1$\times$1$\times$1) phase, $t_{v}$ and $t_{c}$ the amplitudes of the respective dispersions perpendicular to the surface and $m_{v}$, $m_{c}$ the effective masses.

The parameters for equations \ref{dispersions} were derived from photon energy dependent ARPES measurements carried out at the Swiss Light Source on the SIS beamline,
using a Scienta SES-2002 spectrometer with an overall energy resolution better than 10 meV, and an
angular resolution better than 0.5$^{\circ}$. The fit to the data gives for the Se 4p valence band maximum
-20 $\pm$ 10 meV, and for the Ti 3d conduction band a minimum
-40 $\pm$ 5 meV \cite{fits}.
From our measurements
we then find a semimetallic band structure with a negative gap (\textit{i.e.} an overlap)
E$_{G}$=-20 $\pm$ 15 meV for the normal phase of TiSe$_{2}$, in agreement with the literature \cite{PhysRevLett.55.2188}. The dispersions deduced from the ARPES data are shown in fig. \ref{fig1}a) (dashed lines).

Within this model the
one-electron Green's functions of the valence and the conduction
bands were calculated for the excitonic insulator phase. For the
valence band, one obtains
\begin{eqnarray}
G_{v}(\textbf{k},z)=\Big(z-\epsilon_{v}(\textbf{k})-\sum_{\textbf{w}_{i}}\frac{|\Delta|^{2}(\textbf{k},\textbf{w}_{i})}{z-\epsilon_{c}(\textbf{k}+\textbf{w}_{i})}\Big)^{-1} \ .
\end{eqnarray}

This
is a generalized form of the equations of Ref. \cite{PhysRev.158.462} for an arbitrary number
of $\textbf{w}_{i}$. The order parameter $\Delta$
is related to the number of excitons in the condensed state at a given
temperature. For the conduction band, there is a system of equations describing the Green's functions
$G_{c}^{i}$ corresponding to each spanning vector vector $\textbf{w}_{i}$:
\begin{eqnarray}
\big[z-\epsilon_{c}^{i}(\mathbf{k}+\mathbf{w_{i}})\big]G_{c}^{i}(\mathbf{k}+\mathbf{w_{i}},z)
=1+\Delta^{*}(\mathbf{k},\mathbf{w_{i}})\nonumber \\
\times
\sum_{j}\frac{\Delta(\mathbf{k},\mathbf{w_{j}})G_{c}^{j}(\mathbf{k}+\mathbf{w_{j}},z)}{z-\epsilon_{v}(\mathbf{k})}
\end{eqnarray}
This model and the derivation of the Green's functions will be
further described elsewhere \cite{Claude}.

The spectral function calculated along several high-symmetry
directions of the BZ is shown in fig. \ref{fig1}a) for an order parameter $\Delta$=0.05 eV. Its value has been chosen for best agreement with experiment. The color scale shows the spectral weight carried by each band.
For presentation purposes the $\delta$-like peaks of the spectral
function have been broadened by adding a constant 30 meV imaginary part to the self-energy. 
In the normal phase (dashed lines), as previously described we consider a
semimetal with a 20 meV overlap, with bands carrying unity
spectral weight. In the excitonic phase, the band structure is
strongly modified. The first observation is the appearance of new
bands (labeled C1, V2 and C3), backfolded with the spanning vector
$\textbf{w}=\Gamma L$. The C1, V2 and C3 branches are the backfolded replicas of branches C2, V3 and C4 respectively. In this new phase the $\Gamma$ and $L$ points are now equivalent, which means that the excitonic state has a 2$\times$2$\times$2 periodicity of purely electronic origin, as expected from theoretical considerations \cite{PhysRevLett.19.439,PhysRev.158.462}. 
Another effect of exciton condensation is the opening of a gap in the excitation spectrum. This results in a flattening of the valence band near $\Gamma$ in the $\Gamma M$ direction (V1 branch) and in the $A\Gamma$ direction (V3 branch), and also an upward bend of the conduction band near $L$ and $M$ (C2 and C4 branches). It is interesting to
notice that in the vicinity of these two points, the conduction band is split (arrows). This results from the backfolding of the $L$ points onto each other, according to the new periodicity of the excitonic state \cite{PhysRevB.18.2866}. The spectral weight carried by the bands
is shown in fig. \ref{fig1}b).  The largest variations
occur near the $\Gamma$, $L$ and $M$ points, where the band
extrema in the normal phase are close enough for excitons to be
created. Away from these points, the spectral weight decreases in
the backfolded bands (C1, V2, C3) and increases in the others.  
The intensity of the V1 branch, for example, decreases by a factor of 2 when approaching $\Gamma$, whereas the backfolded C1 branch shows the opposite behaviour.
Such a large transfer of spectral weight into the backfolded bands is a very uncommon and striking feature. Indeed, in most
compounds with competing potentials (CDW systems, vicinal surfaces,...), the backfolded bands carry an
extremely small spectral weight \cite{didiot:081404,battaglia:195114,J.Voit10202000}. 
In these systems the backfolding results mainly from the influence of the modified lattice on the electron gas, and the weight transfer is related to the strength of the new crystal potential component. Here, the case of the excitonic insulator is completely different, as the backfolding is an intrinsic property of the excitonic state. The large transfer of spectral weight is then a purely electronic effect, and turns out to be a characteristic feature of the excitonic insulator phase.

\begin{figure}
\begin{center}
\scalebox{0.48}{\includegraphics{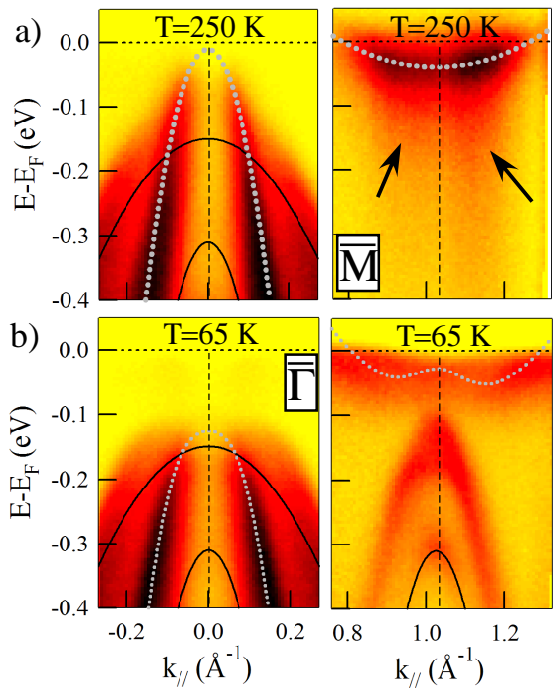}}
\caption{\label{fig2}: ARPES spectra of 1$T$-TiSe$_{2}$ for a) the normal and b) the low temperature phase. Thick dotted lines are
parabolic fits to the bands in the normal phase and thin dotted lines are guides to the eye for the CDW phase. Fine
lines follow the dispersion of the 4p sidebands (see text).}
\end{center}
\end{figure}

Fig. \ref{fig2} shows ARPES spectra recorded at a photon energy h$\nu$=31
eV as a function of temperature.
At this photon energy, the normal emission spectra correspond to
states located close to the $\Gamma$ point. For the sake of simplicity the description is in terms of the surface BZ high-symmetry points $\bar{\Gamma}$ and $\bar{M}$. 
The 250 K spectra exhibit the three Se
4p-derived bands at $\bar{\Gamma}$ and the Ti 3d-derived band at
$\bar{M}$ widely described in the literature
\cite{PhysRevLett.88.226402,PhysRevB.65.235101,PhysRevB.61.16213}.
The thick dotted lines (white) are fits by equation \ref{dispersions}, giving
for the topmost 4p band a maximum energy
of -20 $\pm$ 10 meV, and for the Ti 3d a minimum
energy of -40 $\pm$ 5 meV. The small overlap E$_{G}$=-20 $\pm$ 15 meV in the normal phase is consistent with the excitonic insulator scenario, as the exciton binding energy is expected to be close to that value. \cite{PhysRevLett.88.226402,PhysRevB.61.16213}. 
The position of both band maxima in the occupied states is most probably due to a slight Ti
overdoping of our samples \cite{PhysRevB.14.4321}. In our
case, a transition temperature of 180 $\pm$ 10 K was found from
different ARPES and scanning tunneling microscopy measurements, indicating a Ti doping of less than 1 \%. 
On the 250 K spectrum at $\bar{\Gamma}$, the intensity is
low near normal emission. This reduced
intensity and the residual intensity at $\bar{M}$ around
150 meV binding energy (arrows) 
may arise 
from exciton
fluctuations (see reduction of spectral weight near $\Gamma$ in the V1 branch in fig. 1b). 
Matrix elements do not appear to play a role as the
intensity variation only depends very slightly on photon energy and polarization. 
In the 65 K data (fig. \ref{fig2}b)), the
topmost 4p band flattens near $\bar{\Gamma}$ and shifts to higher binding energy by about 100
meV (thin white, dotted line). This shift is accompanied by a larger decrease of the
spectral weight near the top of the band. The two other bands (fine black lines) are only slightly shifted and do not appear to participate in the transition. In the $\bar{M}$
spectrum strong backfolded valence bands can be seen, and the
conduction band bends upwards, leading to a maximum intensity
located about 0.25 $\AA^{-1}$ from $\bar{M}$ (thin white dotted line). This
observation is in agreement with Kidd \textit{et al.}
\cite{PhysRevLett.88.226402}, although in their case the
conduction band was unoccupied in the normal phase.  

\begin{figure}
\begin{center}
\scalebox{0.48}{\includegraphics{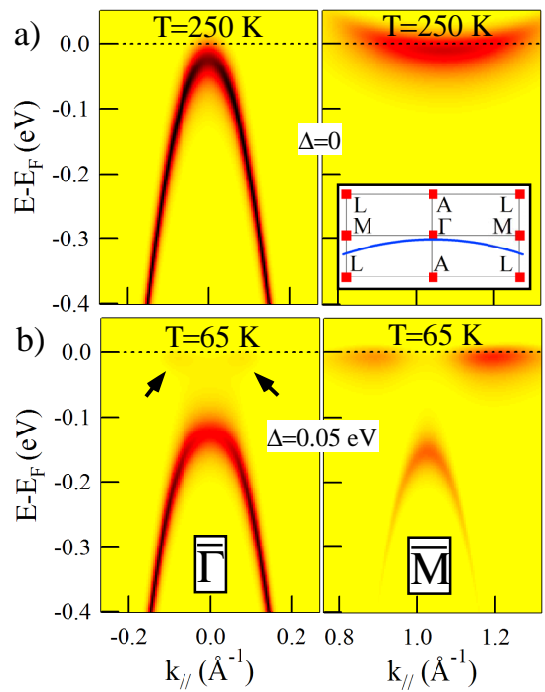}}
\caption{\label{fig3}: Theoretical spectral function of 1$T$-TiSe$_{2}$, calculated along the path given by the
free electron
final state approximation shown in the inset. a) normal state and b) low temperature phase (see text).
}
\end{center}
\end{figure}

The calculated spectral functions corresponding to the data of fig. \ref{fig2}
are shown in fig. \ref{fig3}, using the free-electron final state approximation with a 10 eV inner potential and a 4.6 eV work function (see inset). The effect of temperature was taken into account via the order parameter and the Fermi function. Only the topmost valence band was considered, as the other two are practically not influenced by the transition (see above, fig. 2).  
The behavior of this band is extremely  well reproduced by the calculation. In the 65 K calculation 
the valence band is flattened near $\bar{\Gamma}$, and the spectral weight at this point is reduced to 44 $\%$, close to the experimental value of 35 $\%$. The agreement is very satisfying, considering that the calculation takes into account only the lowest excitonic state. 
The experimental features appear broader than in the calculation, but at finite temperatures one may expect the existence of excitons with non-zero momentum, leading to a spread of spectral weight away from the high-symmetry points. 

In the near-$\bar{M}$ spectral function, the backfolded valence band is strongly present in the 65 K calculation, with comparable spectral weight as at $\bar{\Gamma}$ and as the conduction band at $\bar{M}$. The
conduction band maximum intensity is located away from $\bar{M}$ as in the experiment. The small perpendicular dispersion of the free-electron final
state causes an asymmetry of the intensity of the conduction band
on each side of $\bar{M}$, which is also visible in fig.
\ref{fig2}. In our
calculation, as opposed to the ARPES spectra, the conduction band is unoccupied and only the occupied tail of the peaks is visible. This difference
may be simply due to the final state approximation used in the
calculation, a slight shift of the chemical potential due to the transition, or to atomic displacements 
that would shift the conduction band \cite{PhysRevB.65.235101,PhysRevLett.88.226402,WhangboMyungHwan_ja00050a044}. Such atomic displacements, in terms of a band Jahn-Teller effect, were suggested as a driving force for the transition. However, the key point is that, although the lattice distortion may shift the conduction band, the very small atomic 
displacements ($\approx$ 0.02 $\AA$ \cite{PhysRevB.14.4321}) in 1$T$-TiSe$_{2}$ are expected to lead to a negligable spectral weight in the backfolded bands \cite{J.Voit10202000}. 
As an example, 1$T$-TaS$_2$, another CDW compound known for very large atomic displacements \cite{Spijkerman1997} (of order $>$  0.1 $\AA$) introduces hardly detectable backfolding of spectral weight in ARPES. Clearly, an electronic origin is necessary for obtaining such strong backfolding in the presence of such small atomic displacements.
Therefore, our results allow to rule out a Jahn-Teller effect as the driving force for the transition of TiSe$_{2}$. 

\begin{figure}
\begin{center}
\scalebox{0.48}{\includegraphics{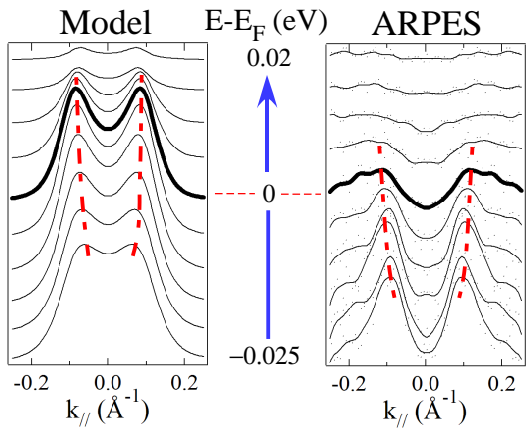}}
\caption{\label{fig4}: Near-E$_{F}$ constant energy cuts in the vicinity of the
$\Gamma$ point. The theoretical data correspond to fig. \ref{fig3}b
and the ARPES data are taken from the low-temperature data of fig.
\ref{fig2}b.}
\end{center}
\end{figure}

Furthermore, the ARPES spectra also show evidence for the backfolded conduction band
 at the $\bar{\Gamma}$ point. Fig. \ref{fig4} shows
constant energy cuts around the Fermi energy, taken from the data
of fig. \ref{fig2}b and \ref{fig3}b (arrows). In the ARPES data two
slightly dispersive peaks, reproduced in the calculation, clearly cross
the Fermi level. These features turn out to be the
populated tail of the backfolded conduction band, whose centroid
is located just above the Fermi level. To our knowledge no
evidence for the backfolding of the conduction band had been put
forward so far.

In summary, by comparing ARPES spectra of 1\textit{T}-TiSe$_{2}$
to theoretical predictions for an excitonic insulator, we have
shown that the superperiodicity of the excitonic state with respect to the lattice results in a very large transfer of spectral weight into backfolded bands. This effect, clearly evidenced by photoemission, turns out to be a characteristic feature of the excitonic insulator phase, thus giving strong  evidence for the existence of this phase in 1\textit{T}-TiSe$_{2}$ and its prominent role in the CDW transition. 

Skillfull technical assistance was provided 
by the workshop and electric engineering team.
This work
was supported by the Fonds National Suisse pour la
Recherche Scientifique through Div. II and MaNEP.


\begin{thebibliography}{10}
\bibitem{fits} 
The fit parameters are : $\epsilon_{v}^{0}$=-0.08$\pm$0.005 eV, $m_{v}$=-0.23$\pm$0.02 $m_{e}$,
where $m_{e}$ is the free electron mass, $t_{v}$=0.06$\pm$0.005 eV ; $\epsilon_{c}^{0}$=-0.01$\pm$0.0025 eV, $m_{c}^{x}$=5.5$\pm$0.2 $m_{e}$, $m_{c}^{y}$=2.2$\pm$0.1 $m_{e}$, $t_{c}$=0.03$\pm$0.0025 eV
\end{thebibliography}
\end{document}